\documentclass[letterpaper]{article}
\usepackage{aaai}
\usepackage{times}
\usepackage{helvet}
\usepackage{courier}
\usepackage{subfigure}
\usepackage[hyphens]{url}
\usepackage{graphicx}

\newcommand{\squishlist}{
\begin{list}
	{$\bullet$} { \setlength{
	\itemsep}{0pt} \setlength{\parsep}{3pt} \setlength{\topsep}{3pt} \setlength{
	\partopsep}{0pt} \setlength{\leftmargin}{1.5em} \setlength{\labelwidth}{1em} \setlength{\labelsep}{0.5em} } }
	
	\newcommand{\squishlisttwo}{
	\begin{list}
		{$\bullet$} { \setlength{
		\itemsep}{0pt} \setlength{\parsep}{0pt} \setlength{\topsep}{0pt} \setlength{
		\partopsep}{0pt} \setlength{\leftmargin}{2em} \setlength{\labelwidth}{1.5em} \setlength{\labelsep}{0.5em} } }
		
		\newcommand{\squishend}{
	\end{list}
	}

\begin{document}
%
\title{The Dark Side of Micro-Task Marketplaces: Characterizing Fiverr and Automatically Detecting Crowdturfing}
\author{Kyumin Lee\\
Utah State University\\
Logan, UT 84322\\
kyumin.lee@usu.edu\\
\And
Steve Webb\\
Georgia Institute of Technology\\
Atlanta, GA 30332\\
steve.webb@gmail.com\\
\And
Hancheng Ge\\
Texas A\&M University\\
College Station, TX 77843\\
hge@cse.tamu.edu\\
}
\maketitle
\begin{abstract}
As human computation on crowdsourcing systems has become popular and powerful for performing tasks, malicious users have started misusing these systems by posting malicious tasks, propagating manipulated contents, and targeting popular web services such as online social networks and search engines. Recently, these malicious users moved to Fiverr, a fast-growing micro-task marketplace, where workers can post crowdturfing tasks (i.e., astroturfing campaigns run by crowd workers) and malicious customers can purchase those tasks for only \$5. In this paper, we present a comprehensive analysis of Fiverr. First, we identify the most popular types of crowdturfing tasks found in this marketplace and conduct case studies for these crowdturfing tasks. Then, we build crowdturfing task detection classifiers to filter these tasks and prevent them from becoming active in the marketplace. Our experimental results show that the proposed classification approach effectively detects crowdturfing tasks, achieving 97.35\% accuracy. Finally, we analyze the real world impact of crowdturfing tasks by purchasing active Fiverr tasks and quantifying their impact on a target site. As part of this analysis, we show that current security systems inadequately detect crowdsourced manipulation, which confirms the necessity of our proposed crowdturfing task detection approach.
\end{abstract}

\section{Introduction}
Crowdsourcing systems are becoming more and more popular because they can quickly accomplish tasks that are difficult for computers but easy for humans. For example, a word document can be summarized and proofread by crowd workers while the document is still being written by its author~\cite{Bernstein:2010}, and missing data in database systems can be populated by crowd workers~\cite{Franklin:2011}. As the popularity of crowdsourcing has increased, various systems have emerged -- from general-purpose crowdsourcing platforms such as Amazon Mechanical Turk, Crowdflower and Fiverr, to specialized systems such as Ushahidi (for crisis information) and Foldit (for protein folding).

These systems offer numerous positive benefits because they efficiently distribute jobs to a workforce of willing individuals.  However, malicious customers and unethical workers have started misusing these systems, spreading malicious URLs in social media, posting fake reviews and ratings, forming artificial grassroots campaigns, and manipulating search engines (e.g., creating numerous backlinks to targeted pages and artificially increasing user traffic). Recently, news media reported that 1,000 crowdturfers -- workers performing crowdturfing tasks on behalf of buyers -- were hired by Vietnamese propaganda officials to post comments that supported the government~\cite{Vietnam:2013}, and the ``Internet water army'' in China created an artificial campaign to advertise an online computer game \cite{abs-1111-4297,Wired:2010}. These types of crowdsourced manipulations reduce the quality of online social media, degrade trust in search engines, manipulate political opinion, and eventually threaten the security and trustworthiness of online web services. Recent studies found that $\sim$90\% of all tasks in crowdsourcing sites were for ``crowdturfing'' -- astroturfing campaigns run by crowd workers on behalf of customers -- \cite{Wang:2012}, and most malicious tasks in crowdsourcing systems target either online social networks (56\%) or search engines (33\%) \cite{Lee13icwsm}.

Unfortunately, very little is known about the properties of crowdturfing tasks, their impact on the web ecosystem, or how to detect and prevent them. Hence, in this paper we are interested in analyzing Fiverr -- a fast growing micro-task marketplace and the 125th most popular site \cite{Alexa:2013} -- to be the first to answer the following questions: what are the most important characteristics of buyers (a.k.a. customers) and sellers (a.k.a. workers)? What types of tasks, including crowdturfing tasks, are available? What sites do crowdturfers target? How much do they earn? Based on this analysis and the corresponding observations, can we automatically detect these crowdturfing tasks? Can we measure the impact of these crowdturfing tasks? Can current security systems in targeted sites adequately detect crowdsourced manipulation?

To answer these questions, we make the following contributions in this paper:

\squishlist
\item First, we collect a large number of active tasks (these are called gigs in Fiverr) from all categories in Fiverr. Then, we analyze the properties of buyers and sellers as well as the types of crowdturfing tasks found in this marketplace. To our knowledge, this is the first study to focus primarily on Fiverr.
\item Second, we conduct a statistical analysis of the properties of crowdturfing and legitimate tasks, and we build a machine learning based crowdturfing task classifier to actively filter out these existing and new malicious tasks, preventing propagation of crowdsourced manipulation to other web sites. To our knowledge this is the first study to detect crowdturfing tasks automatically.
\item Third, we feature case studies of three specific types of crowdturfing tasks: social media targeting gigs, search engine targeting gigs and user traffic targeting gigs.
\item Finally, we purchase active crowdturfing tasks targeting a popular social media site, Twitter, and measure the impact of these tasks to the targeted site. We then test how many crowdsourced manipulations Twitter's security can detect, and confirm the necessity of our proposed crowdturfing detection approach.

\squishend

\section{Background}
Fiverr is a micro-task marketplace where users can buy and sell services, which are called \emph{gigs}. The site has over 1.7 million registered users, and it has listed more than 2 million gigs\footnote{http://blog.Fiverr/2013/08/12/fiverr-community-milestone-two-million-reasons-to-celebrate-iamfiverr/}. As of November 2013, it is the 125th most visited site in the world according to Alexa \cite{Alexa:2013}.

Fiverr gigs do not exist in other e-commerce sites, and some of them are humorous (e.g., ``I will paint a logo on my back'' and ``I will storyboard your script'').  In the marketplace, a \emph{buyer} purchases a gig from a \emph{seller} (the default purchase price is \$5). A user can be a buyer and/or a seller. A buyer can post a review about the gig and the corresponding seller. Each seller can be promoted to a 1st level seller, a 2nd level seller, or a top level seller by selling more gigs. Higher level sellers can sell additional features (called ``gig extras'') for a higher price (i.e., more than \$5). For example, one seller offers the following regular gig: ``I will write a high quality 100 to 300 word post,article,etc under 36 hrs free editing for \$5''. For an additional \$10, she will ``make the gig between 600 to 700 words in length'', and for an additional \$20, she will ``make the gig between 800 to 1000 words in length''. By selling these extra gigs, the promoted seller can earn more money. Each user also has a profile page that displays the user's bio, location, reviews, seller level, gig titles (i.e., the titles of registered services), and number of sold gigs.

\begin{figure}
\centering
\includegraphics[width=3.0in]{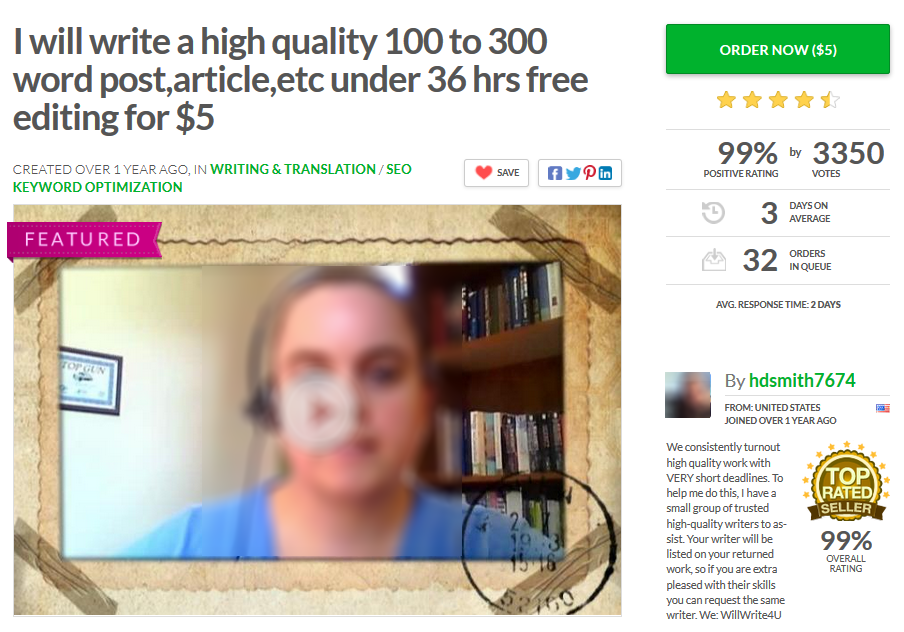}
\caption{An example of a Fiverr gig listing.}
\label{fig:gig}
\end{figure}

Figure~\ref{fig:gig} shows an example of a gig listing on Fiverr. The listing's human-readable URL is http://Fiverr.com/hdsmith7674/write-a-high-quality-100-to-300-word-postarticleetc-under-36-hrs-free-editing, which was automatically created by Fiverr based on the title of the gig. The user name is ``hdsmith7674'', and the user is a top rated seller.

\begin{figure*}
\centering
\includegraphics[width=6.0in]{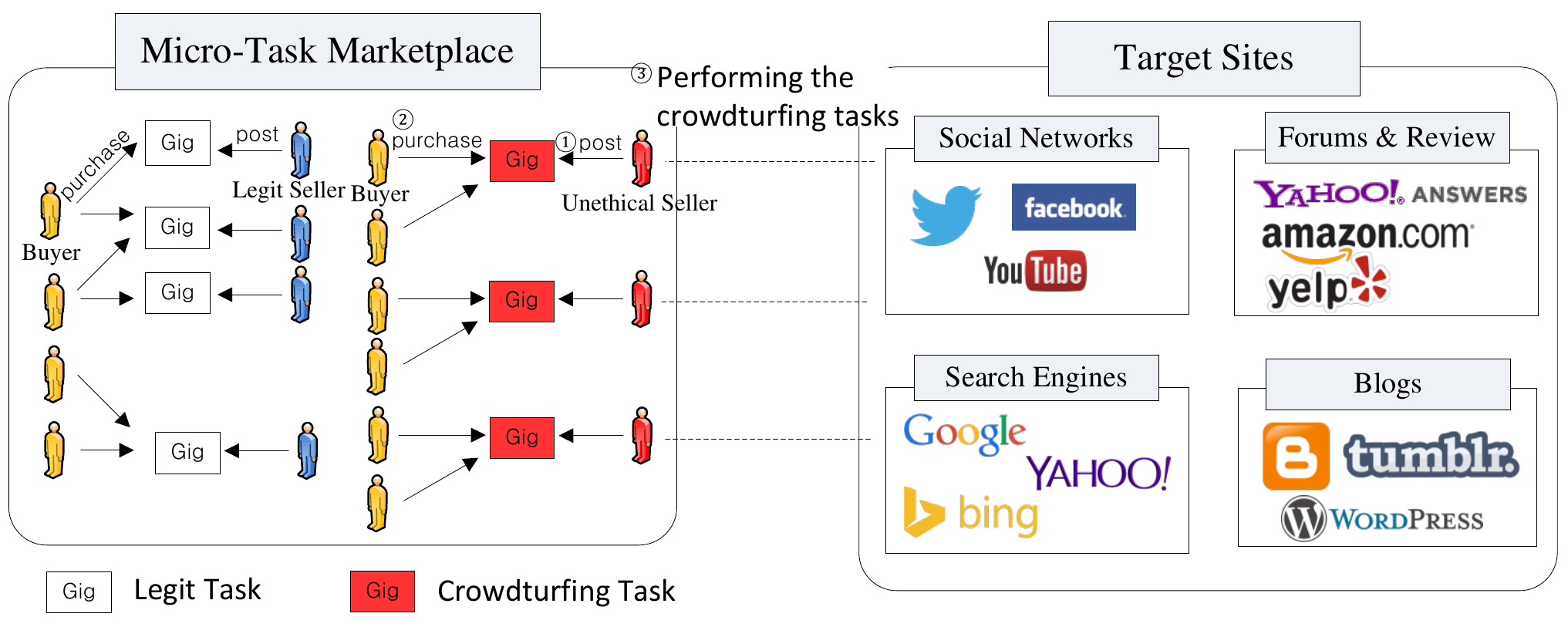}
\caption{The interactions between buyers and legitimate sellers on Fiverr, contrasted with the interactions between buyers and unethical sellers.}
\label{fig:fiverr}
\end{figure*}

Ultimately, there are two types of Fiverr sellers: (1) legitimate sellers and (2) unethical (malicious) sellers, as shown in Figure~\ref{fig:fiverr}. Legitimate sellers post legitimate gigs that do not harm other users or other web sites. Examples of legitimate gigs are ``I will color your logo'' and ``I will sing a punkrock happy birthday''. On the other hand, unethical sellers post crowdturfing gigs on Fiverr that target sites such as online social networks and search engines. Examples of crowdturfing gigs are ``I will provide 2000+ perfect looking twitter followers'' and ``I will create 2,000 Wiki Backlinks''. These gigs are clearly used to manipulate their targeted sites and provide an unfair advantage for their buyers.

\section{Fiverr Characterization}
In this section, we present our data collection methodology. Then, we measure the number of the active Fiverr gig listings and estimate the number of listings that have ever been created. Finally, we analyze the characteristics of Fiverr buyers and sellers.

\subsection{Dataset}
\label{sec:dataset}
To collect gig listings, we built a custom Fiverr crawler. This crawler initially visited the Fiverr homepage and extracted its embedded URLs for gig listings. Then, the crawler visited each of those URLs and extracted new URLs for gig listings using a depth-first search. By doing this process, the crawler accessed and downloaded each gig listing from all of the gig categories between July and August 2013. From each listing, we also extracted the URL of the associated seller and downloaded the corresponding profile. Overall, we collected 89,667 gig listings and 31,021 corresponding user profiles.


%

\subsection{Gig Analysis}
First, we will analyze the gig listings in our dataset and answer relevant questions.

\smallskip
\noindent\textbf{How much data was covered?} We attempted to collect every active gig listing from every gig category in Fiverr. To check how many active listings we collected, we used a sampling approach. When a listing is created, Fiverr internally assigns a sequentially increasing numerical id to the gig. For example, the first created listing received 1 as the id, and the second listing received 2. Using this number scheme, we can access a listing using the following URL format: http://Fiverr.com/[GIG\_NUMERICAL\_ID], which will be redirected to the human-readable URL that is automatically assigned based on the gig's title.

As part of our sampling approach, we sampled 1,000 gigs whose assigned id numbers are between 1,980,000 and 1,980,999 (e.g., http://Fiverr.com/1980000). Then, we checked how many of those gigs are still active because gigs are often paused or deleted. 615 of the 1,000 gigs were still active. Next, we crossreferenced these active listings with our dataset to see how many listings overlapped. Our dataset contained 517 of the 615 active listings, and based on this analysis, we can approximate that our dataset covered 84\% of the active gigs on Fiverr. This analysis also shows that gig listings can become stale quickly due to frequent pauses and deletions.

Initially, we attempted to collect listings using gig id numbers (e.g., http://Fiverr.com/1980000), but Fiverr's Safety Team blocked our computers' IP addresses because accessing the id-based URLs is not officially supported by the site.  To abide by the site's policies, we used the human-readable URLs, and as our sampling approach shows, we still collected a significant number of active Fiverr gig listings.



\begin{figure}
\centering
\includegraphics[width=\linewidth]{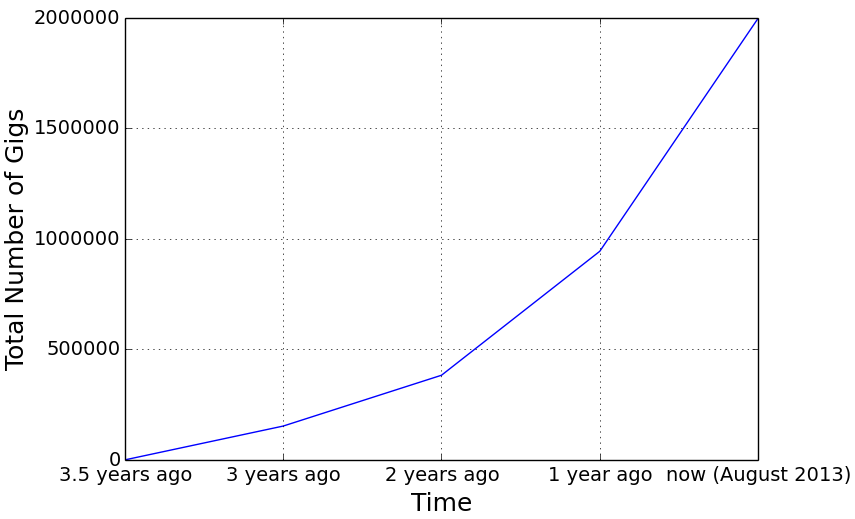}
\caption{Total number of created gigs over time.}
\label{fig:gig-distribution}
\end{figure}

\smallskip
\noindent\textbf{How many gigs have been created over time?} A gig listing contains the gig's numerical id and its creation time, which is displayed as days, months, or years. Based on this information, we can measure how many gigs have been created over time. In Figure~\ref{fig:gig-distribution}, we plotted the approximate total number of gigs that have been created each year. The graph follows the exponential distribution in macro-scale (again, yearly) even though the micro-scaled plot may show us a clearer growth rate. This plot shows that Fiverr has been getting more popular, and in August 2013, the site reached 2 million listed gigs.

\subsection{User Analysis}
Next, we will analyze the characteristics of Fiverr buyers and sellers in the dataset.

\begin{figure*}
\centering
\subfigure[Distribution of sellers in the world map.] 
{
    \label{fig:seller-dist}
    \includegraphics[width=0.45\linewidth]{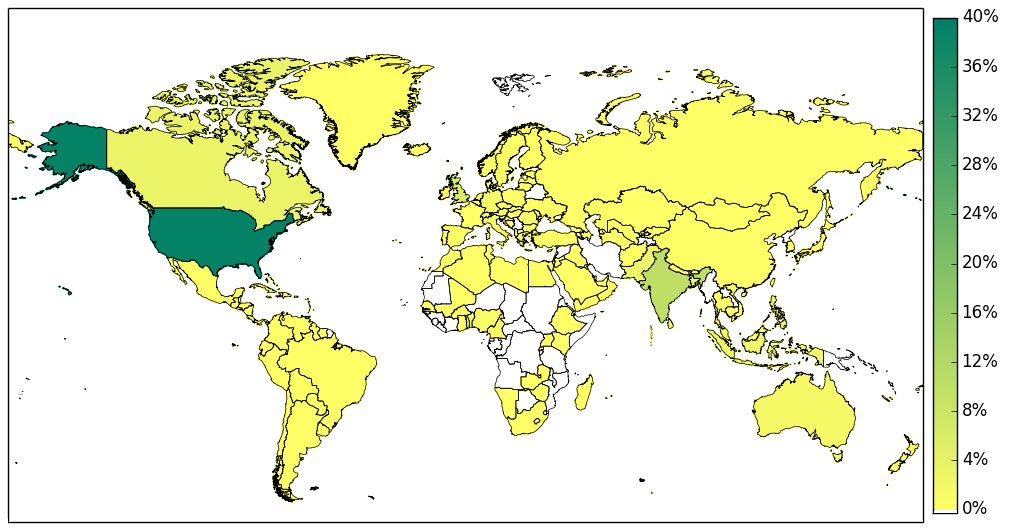}
}
\hspace{0.03cm}
\subfigure[Distribution of buyers in the world map.] 
{
    \label{fig:buyer-dist}
    \includegraphics[width=0.45\linewidth]{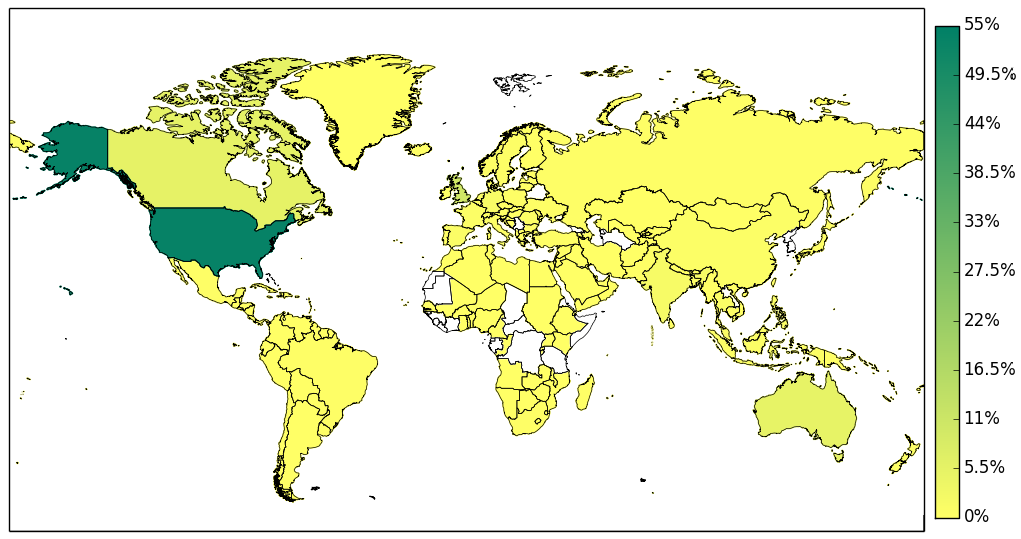}
}
\caption{Distribution of all sellers and buyers in the world map.}
\label{fig:seller-buyer} 
\end{figure*}

\smallskip
\noindent\textbf{Where are sellers from?} Are the sellers distributed all over the world? In previous research, sellers (i.e., workers) in other crowdsourcing sites were usually from developing countries~\cite{Lee13icwsm}. To determine if Fiverr has the same demographics, Figure~\ref{fig:seller-dist} shows the distribution of sellers on the world map. Sellers are from 168 countries, and surprisingly, the largest group of sellers are from the United States (39.4\% of the all sellers), which is very different from other sites. The next largest group of sellers is from India (10.3\%), followed by the United Kingdom (6.2\%), Canada (3.4\%), Pakistan (2.8\%), Bangladesh (2.6\%), Indonesia (2.4\%), Sri Lanka (2.2\%), Philippines (2\%), and Australia (1.6\%). Overall, the majority of sellers (50.6\%) were from the western countries.

%

\begin{table*}[!ht]
\centering
\caption{Top 10 sellers.}
\begin{tabular}{|l|r|r|r|l|c|} \hline
Username & $|$Sold Gigs$|$ & Eared (Minimum) & $|$Gigs$|$ & Gig Category & Crowdturfing\\ \hline
crorkservice & 601,210 & 3,006,050 & 29 & Online Marketing & yes  \\ \hline
dino\_stark & 283,420 & 1,417,100 & 3 & Online Marketing & yes \\ \hline
volarex & 173,030 & 865,150 & 15 & Online Marketing and Advertising & yes\\ \hline
alanletsgo & 171,240 & 856,200 & 29 & Business, Advertising and Online Marketing & yes \\ \hline
portron & 167,945 & 839,725 & 3 & Online Marketing& yes \\ \hline
mikemeth & 149,090 & 745,450 & 19 & Online Marketing, Business, Advertising & yes\\ \hline
actualreviewnet & 125,530 & 627,650 & 6 & Graphics \& Design, Gift and Fun \& Bizarre & no\\ \hline
bestoftwitter & 123,725 & 618,625 & 8 & Online Marketing and Advertising & yes \\ \hline
amazesolutions & 99,890 & 499,450 & 1 & Online Marketing & yes \\ \hline
sarit11 & 99,320 & 496,600 & 2 & Online Marketing & yes  \\ \hline
\end{tabular}
\label{table:top-sellers}
\end{table*}

\smallskip
\noindent\textbf{What is Fiverr's market size?} We analyzed the distribution of purchased gigs in our dataset and found that a total of 4,335,253 gigs were purchased from the 89,667 unique listings. In other words, the 31,021 users in our dataset sold more than 4.3 million gigs and earned at least \$21.6 million, assuming each gig's price was \$5. Since some gigs cost more than \$5 (due to gig extras), the total gig-related revenue is probably even higher. Obviously, Fiverr is a huge marketplace, but where are the buyers coming from? Figure~\ref{fig:buyer-dist} shows the distribution of sold gigs on the world map. Gigs were bought from all over the world (208 total countries), and the largest number of gigs (53.6\% of the 4,335,253 sold gigs) were purchased by buyers in the United States. The next most frequent buyers are the United Kingdom (10.3\%), followed by Canada (5.5\%), Australia (5.2\%), and India (1.7\%). Based on this analysis, the majority of the gigs were purchased by the western countries.


\smallskip
\noindent\textbf{Who are the top sellers?} The top 10 sellers are listed in Table~\ref{table:top-sellers}. Amazingly, one seller (crorkservice) has sold 601,210 gigs and earned at least \$3 million over the past 2 years. In other words, one user from Moldova has earned at least \$1.5 million/year, which is orders of magnitude larger than \$2,070, the GNI (Gross National Income) per capita of Moldova~\cite{worldbank}. Even the 10th highest seller has earned almost \$500,000. Another interesting observation is that 9 of the top 10 sellers have had multiple gigs that were categorized as online marketing, advertising, or business. The most popular category of these gigs was online marketing.

We carefully investigated the top sellers' gig descriptions to identify which gigs they offered and sold to buyers. Gigs provided by the top sellers (except actualreviewnet) are all crowdturfing tasks, which require sellers to manipulate a web page's PageRank score, artificially propagate a message through a social network, or artificially add friends to a social networking account. This observation indicates that despite the positive aspects of Fiverr, some sellers and buyers have abused the micro-task marketplace, and these crowdturfing tasks have become the most popular gigs. These crowdturfing tasks threaten the entire web ecosystem because they degrade the trustworthiness of information. Other researchers have raised similar concerns about crowdturfing problems and concluded that these artificial manipulations should be detected and prevented~\cite{Wang:2012,Lee13icwsm}. However, previous work has not studied how to detect these tasks. For the remainder of this paper, we will analyze and detect these crowdturfing tasks in Fiverr.


\section{Analyzing and Detecting Crowdturfing Gigs}
In the previous section, we observed that top sellers have earned millions of dollars by selling crowdturfing gigs. Based on this observation, we now turn our attention to studying these crowdturfing gigs in detail and automatically detect them.

\subsection{Data Labeling and 3 Types of Crowdturfing Gigs}
To understand what percentage of gigs in our dataset are associated with crowdturfing, we randomly selected 1,550 out of the 89,667 gigs and labeled them as a legitimate or crowdturfing task. Table~\ref{table:labeled-data} presents the labeled distribution of gigs across 12 top level gig categories predefined by Fiverr. 121 of the 1,550 gigs (6\%) were crowdturfing tasks, which is a significant percentage of the micro-task marketplace. Among these crowdturfing tasks, most of them were categorized as online marketing. In fact, 55.3\% of all online marketing gigs in the sample data were crowdturfing tasks.

\begin{table}
\centering
\caption{Labeled data of randomly selected 1,550 gigs.}
\small
\begin{tabular}{|l|r|r|r|} \hline
Category & $|$Gigs$|$ & $|$Crowdturfing$|$ & Crowdtufing\%\\ \hline
Advertising & 99 & 4 & 4\% \\ \hline
Business & 51 & 1 & 2\% \\ \hline
Fun\&Bizarre & 81 & 0 & 0\% \\ \hline
Gifts & 67 & 0 & 0\%  \\ \hline
Graphics\&Design & 347 & 1 & 0.3\%  \\ \hline
Lifestyle & 114 & 0 & 0\% \\ \hline
Music\&Audio & 123 & 0 & 0\%  \\ \hline
Online Marketing & 206 & 114 & 55.3\% \\ \hline
Other & 20 & 0 & 0 \\ \hline
Programming... & 84 & 0 & 0 \\ \hline
Video\&Animation & 201 & 0 & 0 \\ \hline
Writing\&Trans... & 157 & 1 & 0.6\% \\ \hline
Total & 1,550 & 121 & 6\% \\ \hline
\end{tabular}
\label{table:labeled-data}
\end{table}

Next, we manually categorized the 121 crowdturfing gigs into three groups: (1) social media targeting gigs, (2) search engine targeting gigs, and (3) user traffic targeting gigs. 65 of the 121 crowdturfing gigs targeted social media sites such as Facebook, Twitter and Youtube. The gig sellers know that buyers want to have more friends or followers on these sites, promote their messages or URLs, and increase the number of views associated with their videos. The buyers expect these manipulation to result in more effective information propagation, higher conversion rates, and positive social signals for their web pages and products.

Another group of gigs (47 of the 121 crowdturfing gigs) targeted search engines by artificially creating backlinks for a targeted site. This is a traditional attack against search engines. However, instead of creating backlinks on their own, the buyers take advantage of sellers to create a large number of backlinks so that the targeted page will receive a higher PageRank score (and have a better chance of ranking at the top of search results). The top seller in Table~\ref{table:top-sellers} (crorkservice) has sold search engine targeting gigs and earned \$3 million with 100\% positive ratings and more than 47,000 positive comments from buyers who purchased the gigs. This fact indicates that the search engine targeting gigs are popular and profitable.

The last gig group (9 of the 121 crowdturfing gigs) claimed to pass user traffic to a targeted site. Sellers in this group know that buyers want to generate user traffic (visitors) for a pre-selected web site or web page. With higher traffic, the buyers hope to abuse Google AdSense, which provides advertisements on each buyer's web page, when the visitors click the advertisements. Another goal of purchasing these traffic gigs is for the visitors to purchase products from the pre-selected page.

To this point, we have analyzed the labeled crowdturfing gigs and identified monetization as the primary motivation for purchasing these gigs. By abusing the web ecosystem with crowd-based manipulations, buyers attempt to maximize their profits. In the next section, we will develop an approach to detect these crowdturfing gigs automatically.

\subsection{Detecting Crowdturfing Gigs}
\label{sec:gigdetection}
Automatically detecting crowdturfing gigs is an important task because it allows us to remove the gigs before buyers can purchase them, and eventually, it will allow us to prohibit sellers from posting these gigs. To detect crowdturfing gigs, we built machine-learned models using the manually labeled 1,550 gig dataset.

The performance of a classifier depends on the quality of features, which have distinguishing power between crowdturfing gigs and legitimate gigs in this context. Our feature set consists of the title of a gig, the gig's description, a top level category, a second level category (each gig is categorized to a top level and then a second level -- e.g., ``online marketing'' as the top level and ``social marketing'' as the second level), ratings associated with a gig, the number of votes for a gig, a gig's longevity, a seller's response time for a gig request, a seller's country, seller longevity, seller level (e.g., top level seller or 2nd level seller), a world domination rate (the number of countries where buyers of the gig were from, divided by the total number of countries), and distribution of buyers by country (e.g., entropy and standard deviation). For the title and job description of a gig, we converted these texts into bag-of-word models in which each distinct word becomes a feature. We also used \emph{tf-idf} to measure values for these text features.


To understand which feature has distinguishing power between crowdturfing gigs and legitimate gigs, we measured the chi-square of the features. The most interesting features among the top features, based on chi-square, are category features (top level and second level), a world domination rate, and bag-of-words features such as ``link'', ``backlink'', ``follow'', ``twitter'', ``rank'', ``traffic'', and ``bookmark''.

Since we don't know which machine learning algorithm (or classifier) would perform best in this domain, we tried over 30 machine learning algorithms such as Naive Bayes, Support Vector Machine (SVM), and tree-based algorithms by using the Weka machine learning toolkit with default values for all parameters~\cite{witten:weka}. We used 10-fold cross-validation, which means the dataset containing 1,550 gigs was divided into 10 sub-samples. For a given classification experiment using a single classifier, each sub-sample becomes a testing set, and the other 9 sub-samples become a training set.  We completed a classification experiment for each of the 10 pairs of training and testing sets, and we averaged the 10 classification results. We repeated this process for each machine learning algorithm.

	\begin{table}
        \centering \caption{Confusion matrix}
		\begin{tabular}
			{cc|cc} & & \multicolumn{2}{c}{Predicted}\\
			& & Crowdturfing & Legitimate\\
			\hline Actual & Crowdturfing Gig & ${a}$ & ${b}$\\
			& Legit Gig & ${c}$ & ${d}$\\
		\end{tabular}
		\label{table:confusion-matrix}
	\end{table}
	
We compute precision, recall, F-measure, accuracy, false positive rate (FPR) and false negative rate (FNR) as metrics to evaluate our classifiers. In the confusion matrix, Table~\ref{table:confusion-matrix}, ${a}$ represents the number of correctly classified crowdturfing gigs, ${b}$ (called FNs) represents the number of crowdturfing gigs misclassified as legitimate gigs, ${c}$ (called FPs) represents the number of legitimate gigs misclassified as crowdturfing gigs, and ${d}$ represents the number of correctly classified legitimate gigs. The precision (P) of the crowdturfing gig class is ${a / (a + c)}$ in the table. The recall (R) of the crowdturfing gig is ${a / (a + b)}$. F$_{1}$ measure of the crowdturfing gig class is ${2PR/ (P + R)}$. The accuracy means the fraction of correct classifications and is ${(a + d) / (a + b + c + d)}$.

Overall, SVM outperformed the other classification algorithms. Its classification result is shown in Table~\ref{table:boosting-bagging}. It achieved 97.35\% accuracy, 0.974 F$_{1}$, 0.008 FPR, and 0.248 FNR. This positive result shows that our classification approach works well and that it is possible to automatically detect crowdturfing gigs.
\begin{table}
	[!ht] \centering \caption{SVM-based classification result}
	\begin{tabular}
		{|c|c|c|c|c|c|} \hline
		
        Accuracy & F$_{1}$ & FPR & FNR \\ \hline
		97.35\% & 0.974 & 0.008 & 0.248 \\ \hline
	\end{tabular}
	\label{table:boosting-bagging}
\end{table}

\section{Detecting Crowdturfing Gigs in the Wild and Case Studies}
In this section, we apply our classification approach to a large dataset to find new crowdturfing gigs and conduct case studies of the crowdturfing gigs in detail.

\subsection{Newly Detected Crowdturfing Gigs}
In this study, we detect crowdturfing gigs in the wild, analyze newly detected crowdturfing gigs, and categorize each crowdturfing gig to one of the three crowdturfing types (social media targeting gig, search engine targeting gig, or user traffic targeting gig) revealed in the previous section.

First, we trained our SVM-based classifier with the 1,550 labeled gigs, using the same features as the previous experiment in the previous section. However, unlike the previous experiment, we used all 1,550 gigs as the training set. Since we used the 1,550 gigs for training purposes, we removed those gigs (and 299 other gigs associated with the users that posted the 1,550 gigs) from the large dataset containing 89,667 gigs. After this filtering, the remaining 87,818 gigs were used as the testing set.

We built the SVM-based classifier with the training set and predicted class labels of the gigs in the testing set. 19,904 of the 87,818 gigs were predicted as crowdturfing gigs. Since this classification approach was evaluated in the previous section and achieved high accuracy with a small number of misclassifications for legitimate gigs, almost all of these 19,904 gigs should be real crowdturfing gigs. To make verify this conclusion, we manually scanned the titles of all of these gigs and confirmed that our approach worked well. Here are some examples of these gig titles: ``I will 100+ Canada real facebook likes just within 1 day for \$5'', ``I will send 5,000 USA only traffic to your website/blog for \$5'', and ``I will create 1000 BACKLINKS guaranteed + bonus for \$5''.

\begin{figure}
\centering
\includegraphics[width=\linewidth]{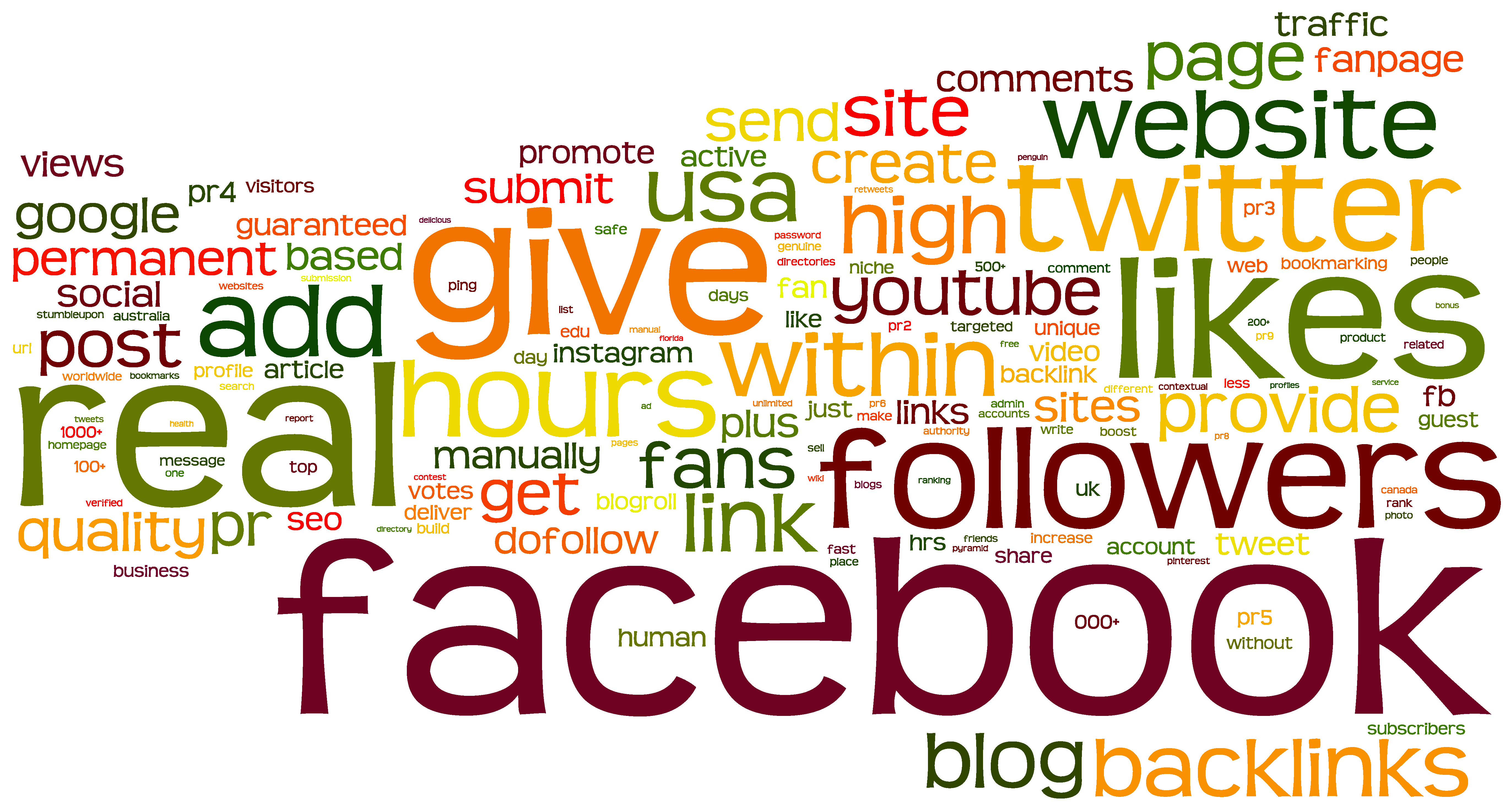}
\caption{Word cloud of crowdturfing gigs.}
\label{fig:word-distribution}
\end{figure}

To understand and visualize what terms crowdturfing gigs often contain, we generated a word cloud of titles for these 19,904 crowdturfing gigs. First, we extracted the titles of the gigs and tokenized them to generate unigrams. Then, we removed stop words. Figure~\ref{fig:word-distribution} shows the word cloud of crowdturfing gigs. The most popular terms are online social network names (e.g., Facebook, Twitter, and YouTube), targeted goals for the online social networks (e.g., likes and followers), and search engine related terms (e.g., backlinks, website, and Google). This word cloud also helps confirm that our classifier accurately identified crowdturfing gigs.

\begin{figure*}
\centering
\subfigure[Sellers.] 
{
    \label{fig:sellers-country}
    \includegraphics[width=0.4\linewidth, height=2.2in]{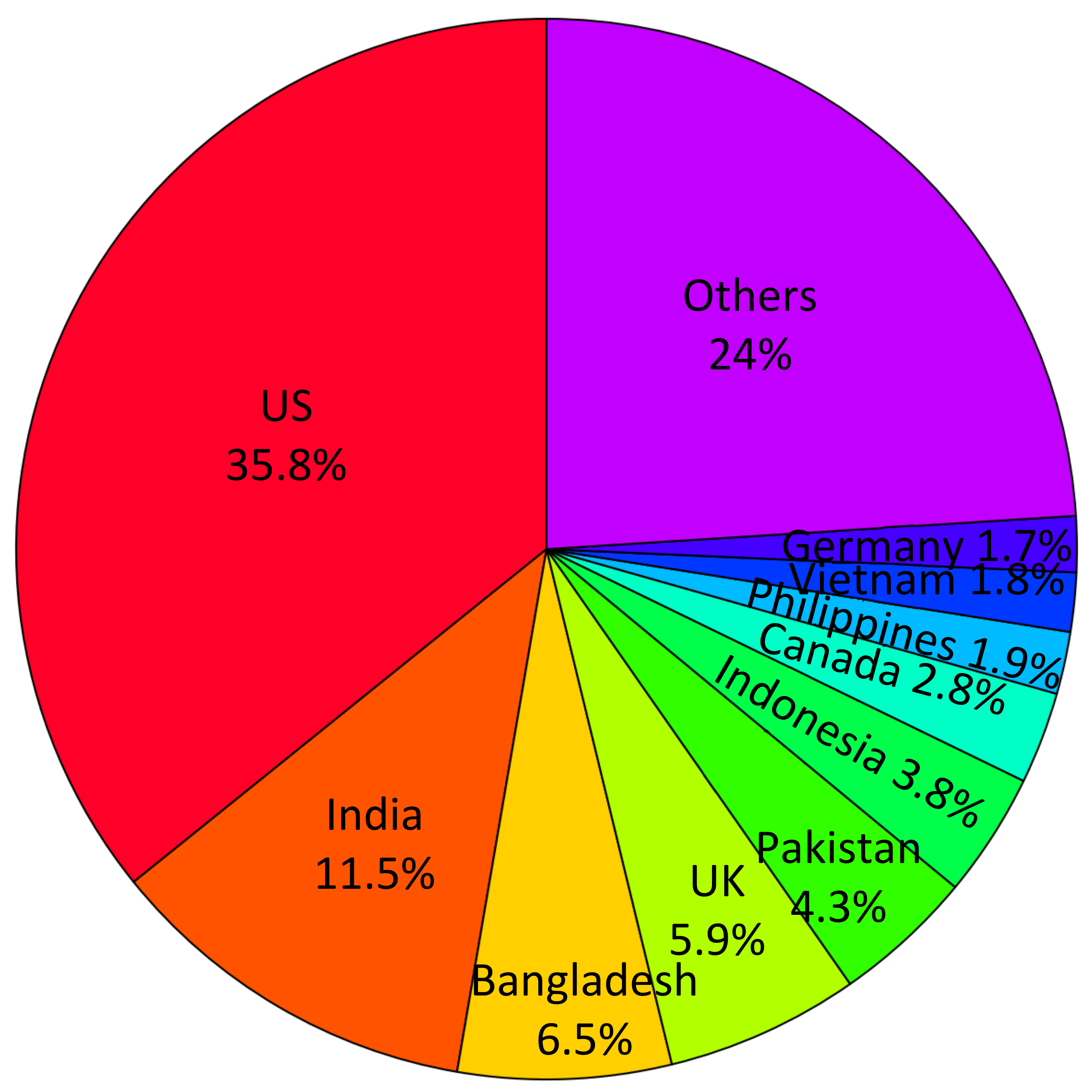}
}
\subfigure[Buyers.] 
{
    \label{fig:buyers-country}
    \includegraphics[width=0.5\linewidth, height=2.2in]{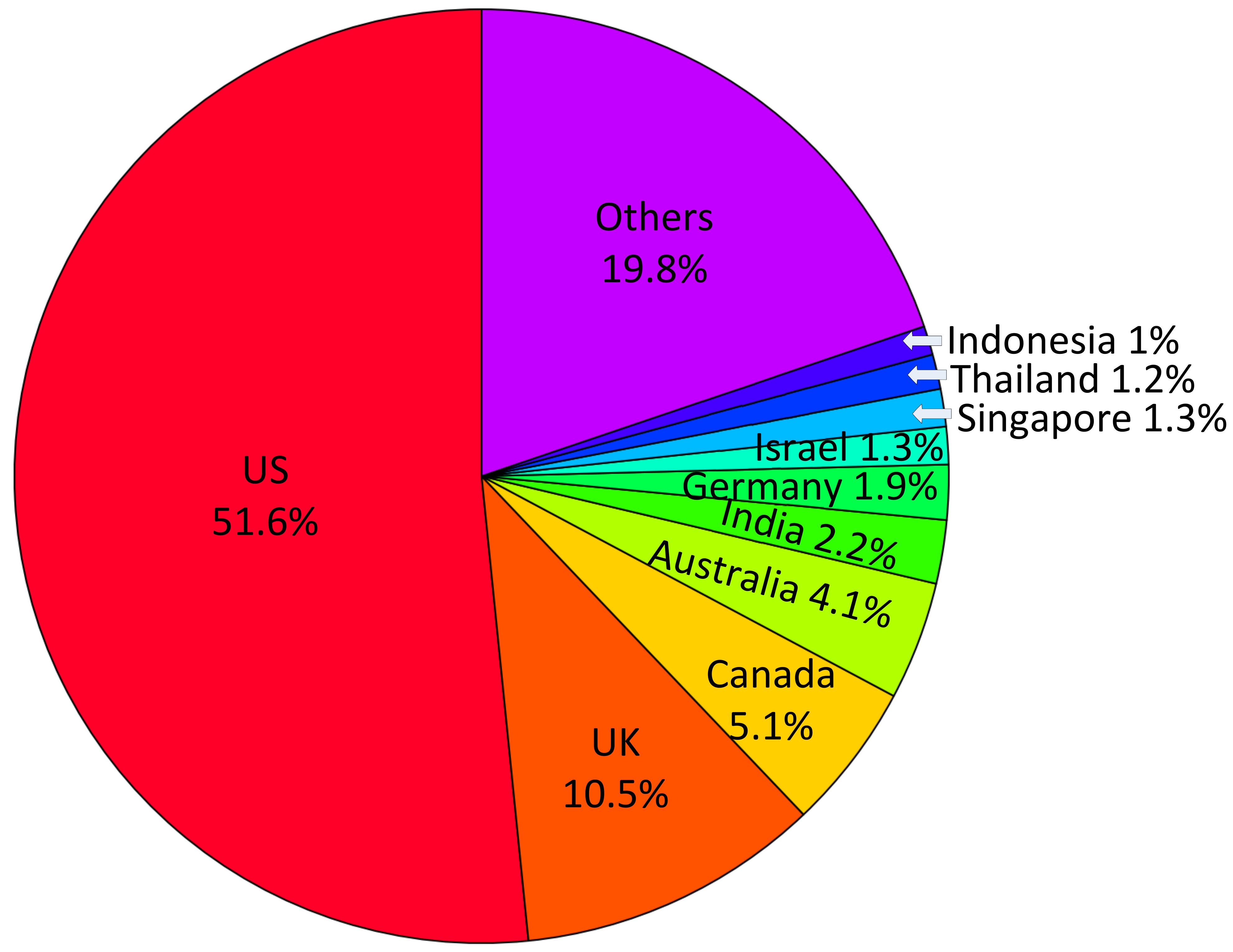}
}
\caption{Top 10 countries of sellers and buyers in crowdturfing gigs.}
\label{fig:top10-countries} 
\end{figure*}

Next, we are interested in analyzing the top 10 countries of buyers and sellers in the crowdturfing gigs. Can we identify different country distributions compared with the distributions of the overall Fiverr sellers and buyers shown in Figure~\ref{fig:seller-buyer}? Are country distributions of sellers and buyers in the crowdsourcing gigs in Fiverr different from distribution of users in other crowdsourcing sites? Interestingly, the most frequent sellers of the crowdturfing gigs in Figure~\ref{fig:sellers-country} were from the United States (35.8\%), following a similar distribution as the overall Fiverr sellers. This distribution is very different from another research result~\cite{Lee13icwsm}, in which the most frequent sellers (called ``workers'' in that research) in another crowdsourcing site, Microworkers.com, were from Bangladesh. This observation might imply that Fiverr is more attractive than Microworkers.com for U.S. residents since selling a gig on Fiverr gives them higher profits (each gig costs at least \$5 but only 50 cents at Microworkers.com). The country distribution for buyers of the crowdturfing gigs in Figure~\ref{fig:buyers-country} is similar with the previous research result~\cite{Lee13icwsm}, in which the majority of buyers (called ``requesters'' in that research) were from English-speaking countries. This is also consistent with the distribution of the overall Fiverr buyers. Based on this analysis, we conclude that the majority of buyers and sellers of the crowdturfing gigs were from the U.S. and other western countries, and these gigs targeted major web sites such as social media sites and search engines.


%


\subsection{Case Studies of 3 Types of Crowdturfing Gigs}
From the previous section, the classifier detected 19,904 crowdturfing gigs. In this section, we classify these 19,904 gigs into the three crowdturfing gig groups in order to feature case studies for the three groups in detail. To further classify the 19,904 gigs into three crowdturfing groups, we built another classifier that was trained using the 121 crowdturfing gigs (used in the previous section), consisting of 65 social media targeting gigs, 47 search engine targeting gigs, and 9 user traffic targeting gigs. The classifier classified the 19,904 gigs as 14,065 social media targeting gigs (70.7\%), 5,438 search engine targeting gigs (27.3\%), and 401 user traffic targeting gigs (2\%). We manually verified that these classifications were correct by scanning the titles of the gigs. Next, we will present our case studies for each of the three types of crowdturfing gigs.

\begin{figure}[!ht]
\centering
\includegraphics[width=\linewidth]{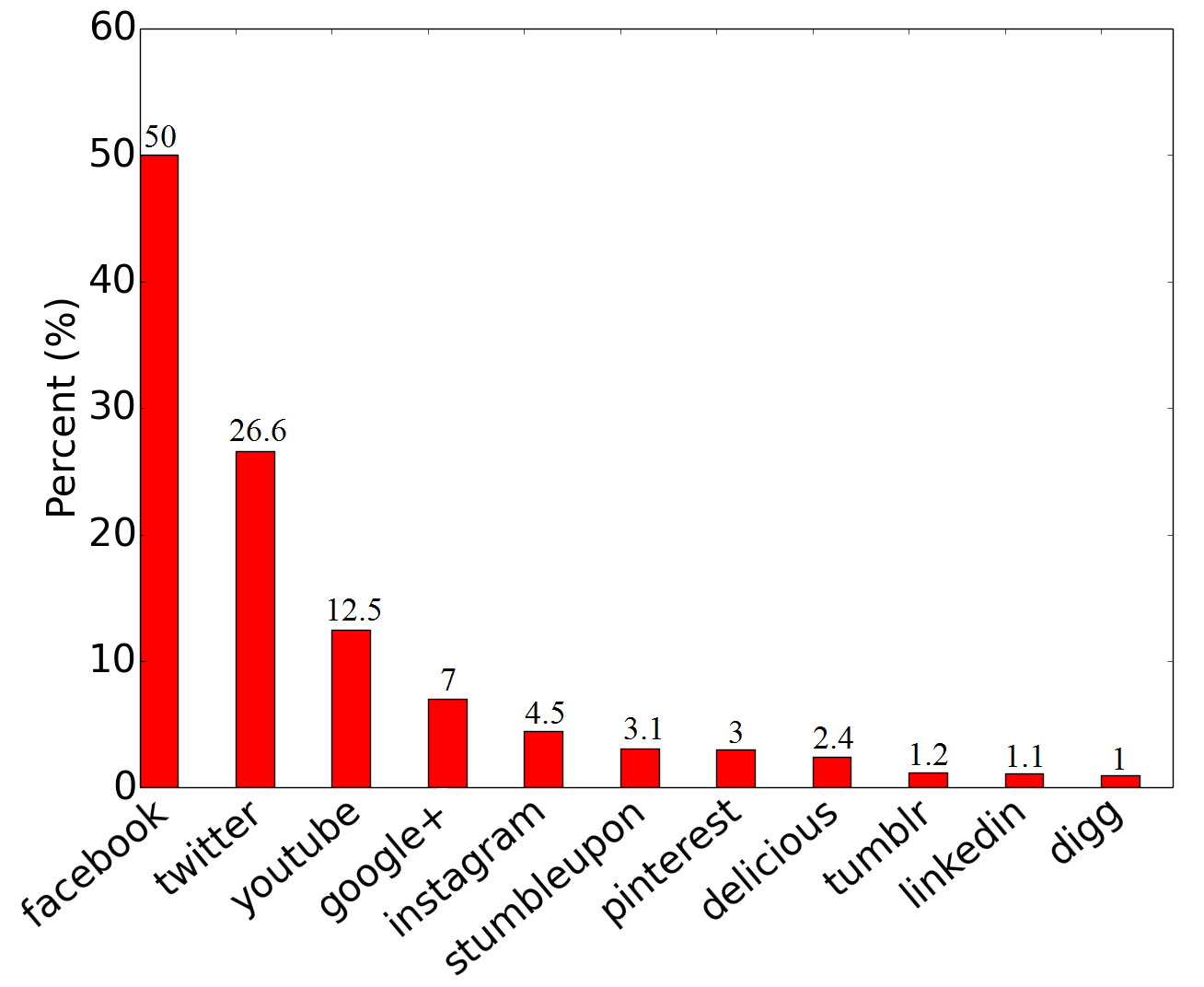}
\caption{Social media sites targeted by the crowdturfing sellers.}
\label{fig:targeted-social-sites}
\end{figure}

\smallskip
\noindent\textbf{Social media targeting gigs.} In Figure~\ref{fig:targeted-social-sites}, we identify the social media sites (including social networking sites) that were targeted the most by the crowdturfing sellers. Overall, most well known social media sites were targeted by the sellers. Among the 14,065 social media targeting gigs, 7,032 (50\%) and 3,744 (26.6\%) gigs targeted Facebook and Twitter, respectively. Other popular social media sites such as Youtube, Google+, and Instagram were also targeted. Some sellers targeted multiple social media sites in a single crowdturfing gig. Example titles for these social media targeting gigs are ``I will deliver 100+ real fb likes from france to you facebook fanpage for \$5'' and ``I will provide 2000+ perfect looking twitter followers without password in 24 hours for \$5''.

\smallskip
\noindent\textbf{Search engine targeting gigs.} People operating a company always have a desire for their web site to be highly ranked in search results generated by search engines such as Google and Bing. The web site's rank order affects the site's profit since web surfers (users) usually only click the top results, and click-through rates of the top pages decline exponentially from \#1 to \#10 positions in a search result~\cite{search}. One popular way to boost the ranking of a web site is to get links from other web sites because search engines measure a web site's importance based on its link structure. If a web site is cited or linked by a well known web site such as cnn.com, the web site will be ranked in a higher position than before. Google's famous ranking algorithm, PageRank, is computed based on the link structure and quality of links. To artificially boost the ranking of web sites, search engine targeting gigs provide a web site linking service. Example titles for these gigs are ``I will build a Linkwheel manually from 12 PR9 Web20 + 500 Wiki Backlinks+Premium Index for \$5'' and ``I will give you a PR5 EDUCATION Nice permanent link on the homepage for \$5''.

\begin{figure}
\centering
\includegraphics[width=\linewidth]{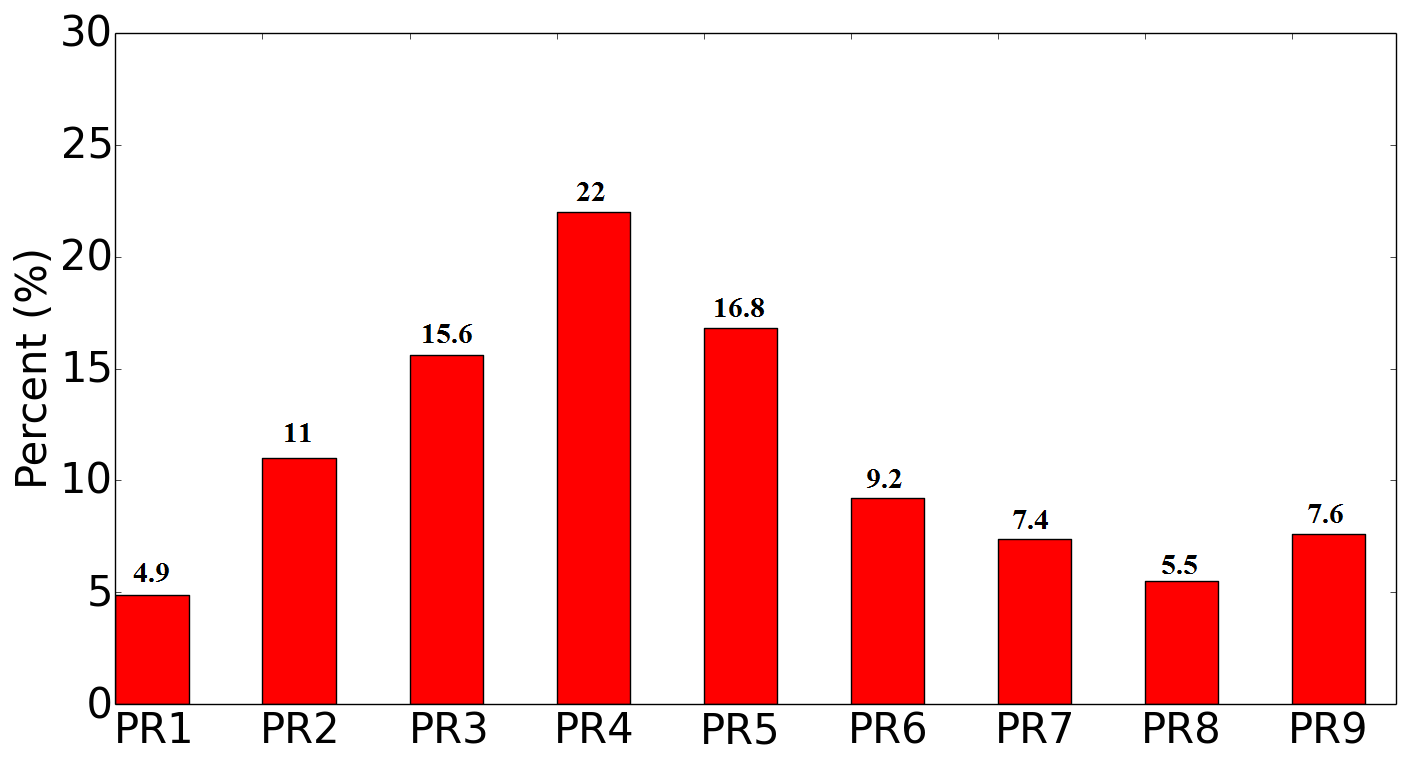}
\caption{PageRank scores of web pages managed by the crowdturfing gig sellers and used to link to a buyer's web page.}
\label{fig:targeted-search-engine}
\end{figure}

As shown in the examples, the sellers of these gigs shared a PageRank score for the web pages that would be used to link to buyers' web sites. PageRank score ranges between 1 and 9, and a higher score means the page's link is more likely to boost the target page's ranking. To understand what types of web pages the sellers provided, we analyzed the titles of the search engine targeting gigs. Specifically, titles of 3,164 (58\%) of the 5,438 search engine targeting gigs explicitly contained a PageRank score of their web pages so we extracted PageRank scores from the titles and grouped the gigs by a PageRank score, as shown in Figure~\ref{fig:targeted-search-engine}. The percentage of web pages between PR1 and PR4 increased from 4.9\% to 22\%. Then, the percentage of web pages between PR5 and PR8 decreased because owning or managing higher PageRank pages is more difficult. Surprisingly, the percentage of PR9 web pages increased. We conjecture that the buyers owning PR9 pages invested time and resources carefully to maintain highly ranked pages because they knew the corresponding gigs would be more popular than others (and much more profitable).

\smallskip
\noindent\textbf{User traffic targeting gigs.} Web site owners want to increase the number of visitors to their sites, called ``user traffic'', to maximize the value of the web site and its revenue. Ultimately, they want these visitors buy products on the site or click advertisements. For example, owners can earn money based on the number of clicks on advertisements supplied from Google AdSense~\cite{adsense}. 401 crowdturfing gigs fulfilled these owners' needs by passing user traffic to buyers' web sites.

\begin{figure}
\centering
\includegraphics[width=\linewidth]{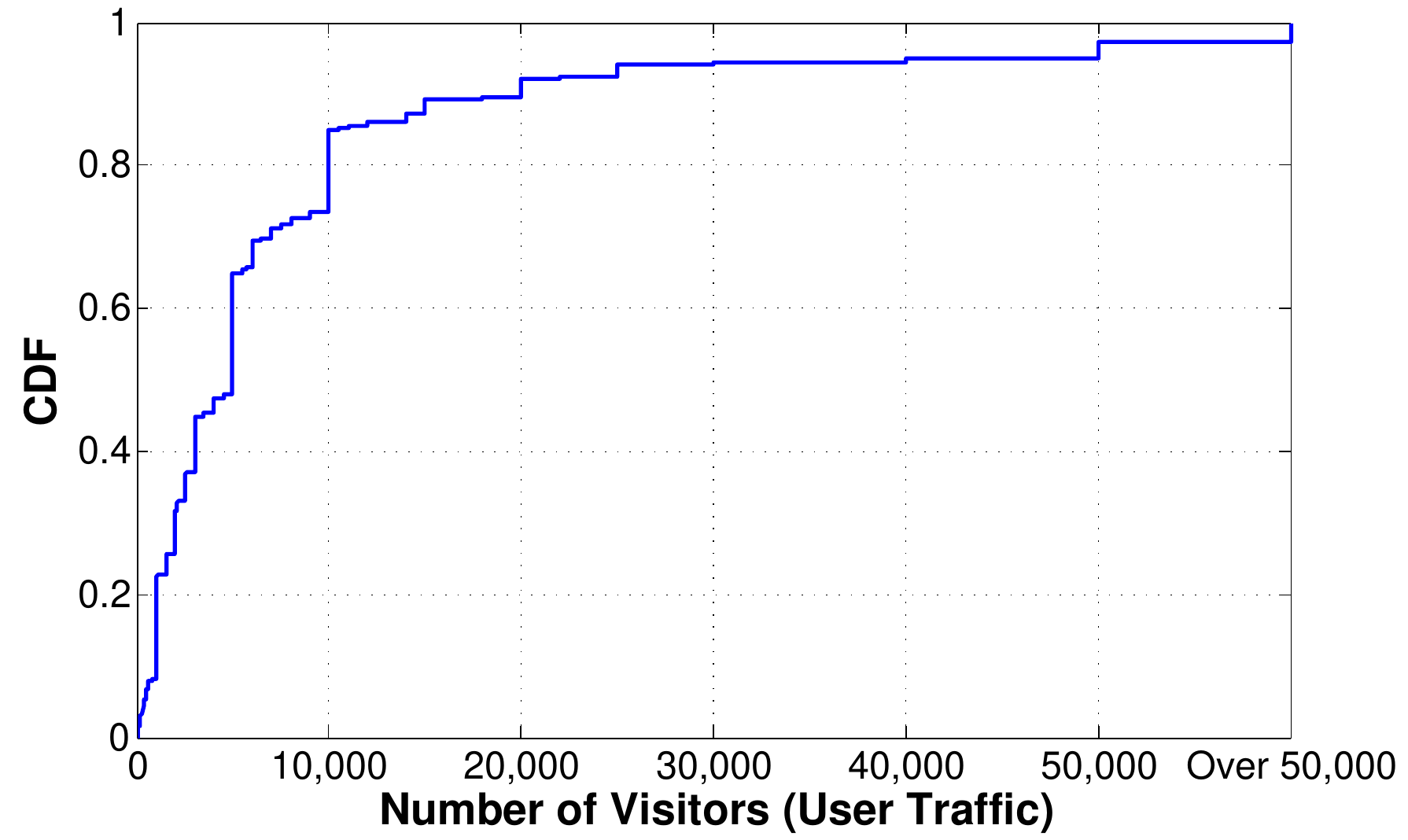}
\caption{The number of visitors (User Traffic) provided by the sellers.}
\label{fig:user-traffic}
\end{figure}

An interesting research question is, ``How many visitors does a seller pass to the destination site of a buyer?'' To answer this question, we analyzed titles of the 401 gigs and extracted the number of visitors by using regular expressions (with manual verification). 307 of the 401 crowdturfing gigs contained a number of expected visitors explicitly in their titles. To visualize these numbers, we plotted the cumulative distribution function (CDF) of the number of promised visitors in Figure~\ref{fig:user-traffic}. While 73\% of sellers guaranteed that they will pass less than 10,000 visitors, the rest of the sellers guaranteed that they will pass 10,000 or more visitors. Even 2.3\% of sellers advertised that they will pass more than 50,000 visitors. Examples of titles for these user traffic targeting gigs are ``I will send 7000+ Adsense Safe Visitors To Your Website/Blog for \$5'' and ``I will send 15000 real human visitors to your website for \$5''. By only paying \$5, the buyers can get a large number of visitors who might buy products or click advertisements on the destination site.

In summary, we identified 19,904 (22.2\%) of the 89,667 gigs as crowdturfing tasks. Among those gigs, 70.7\% targeted social media sites, 27.3\% targeted search engines, and 2\% passed user traffic. The case studies reveal that crowdturfing gigs can be a serious problem to the entire web ecosystem because malicious users can target any popular web service.

\section{Impact of Crowdturfing Gigs}
Thus far, we have studied how to detect crowdturfing gigs and presented case studies for three types of crowdturfing gigs. We have also hypothesized that crowdturfing gigs pose a serious threat, but an obvious question is whether they actually affect to the web ecosystem. To answer this question, we measured the real world impact of crowdturfing gigs. Specifically, we purchased a few crowdturfing gigs targeting Twitter, primarily because Twitter is one of the most targeted social media sites. A common goal of these crowdturfing gigs is to send Twitter followers to a buyer's Twitter account (i.e., artificially following the buyer's Twitter account) to increase the account's influence on Twitter.

To measure the impact of these crowdturfing gigs, we first created five Twitter accounts as the target accounts. Each of the Twitter accounts has a profile photo to pretend to be a human's account, and only one tweet was posted to each account. These accounts did not have any followers, and they did not follow any other accounts to ensure that they are not influential and do not have any friends. The impact of these crowdturfing gigs was measured as a Klout score, which is a numerical value between 0 and 100 that is used to measure a user's influence by Klout\footnote{http://klout.com}. The higher the Klout score is, the more influential the user's Twitter account is. In this setting, the initial Klout scores of our Twitter accounts were all 0.

Then, we selected five gigs that claimed to send followers to a buyer's Twitter account, and we purchased them, using the screen names of our five Twitter accounts. Each of the five gig sellers would pass followers to a specific one of our Twitter accounts (i.e., there was a one seller to one buyer mapping). The five sellers' Fiverr account names and the titles of their five gigs are as follows:
\begin{description}
  \item[spyguyz] I will send you stable 5,000 Twitter FOLLOWERS in 2 days for \$5
  \item[tweet\_retweet] I will instantly add 32000 twitter followers to your twitter account safely \$5
  \item[fiver\_expert] I will add 1000+ Facebook likes Or 5000+ Twitter follower for \$5
  \item[sukmoglea4863] I will add 600 Twitter Followers for you, no admin is required for \$5
  \item[myeasycache] I will add 1000 real twitter followers permanent for \$5
\end{description}

\begin{table}
\centering
\caption{The five gigs' sellers, the number of followers sent by these sellers and the period time took to send all of these followers.}
\begin{tabular}{|l|r|l|} \hline
Seller Name & $|$Sent Followers$|$ & The Period of Time
\\ \hline
spyguyz & 5,502 & within 5 hours \\ \hline
tweet\_retweet & 33,284 & within 47 hours \\ \hline
fiver\_expert & 5,503 & within 1 hour  \\ \hline
sukmoglea4863 & 756 & within 6 hours \\ \hline
myeasycache & 1,315 & within 1 hour \\ \hline
\end{tabular}
\label{table:five-sellers}
\end{table}



%
%
%
%
%
%
%

These sellers advertised sending 5,000, 32,000, 5,000, 600 and 1,000 Twitter followers, respectively. First, we measured how many followers they actually sent us (i.e., do they actually send the promised number of followers?), and then, we identify how quickly they sent the followers. Table~\ref{table:five-sellers} presents the experimental result. Surprisingly, all of the sellers sent a larger number of followers than they originally promised. Even tweet\_retweet sent almost 33,000 followers for just \$5. While tweet\_retweet sent the followers within 47 hours (within 2 days, as the seller promised), the other four sellers sent followers within 6 hours (two of them sent followers within 1 hour). In summary, we were able to get a large number of followers (more than 45,000 followers in total) by paying only \$25, and these followers were sent to us very quickly.

\begin{figure}
\centering
\includegraphics[width=\linewidth]{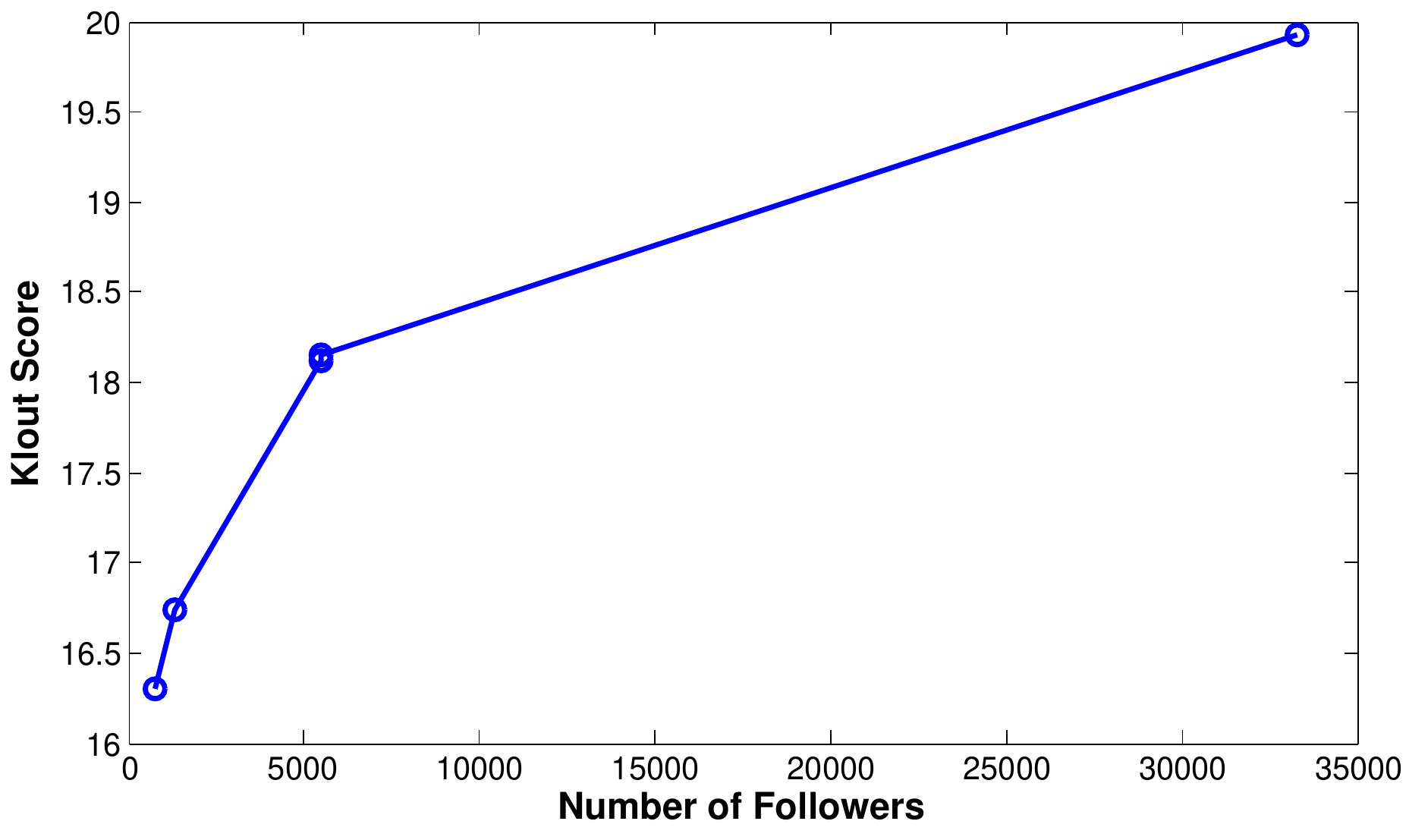}
\caption{Klout scores of our five Twitter accounts were correlated with number of followers of them.}
\label{fig:impact}
\end{figure}

Next, we measured the impact of these artificial Twitter followers by checking the Klout scores for our five Twitter accounts (again, our Twitter accounts' initial Klout scores were 0). Specifically, after our Twitter accounts received the above followers from the Fiverr sellers, we checked their Klout scores to see whether artificially getting followers improved the influence of our accounts. In Klout, the higher a user's Klout score is, the more influential the user is~\cite{klout}. Surprisingly, the Klout scores of our accounts were increased to 18.12, 19.93, 18.15, 16.3 and 16.74, which corresponded to 5,502, 33,284, 5,503, 756 and 1,316 followers. From this experimental result, we learned that an account's Klout score is correlated with its number of followers, as shown in Figure~\ref{fig:impact}. Apparently, getting followers (even artificially) increased the Klout scores of our accounts and made them more influential. In summary, our crowdsourced manipulations had a real world impact on a real system. 

\smallskip
\noindent\textbf{The Followers Suspended By Twitter.}
Another interesting research question is, ``Can current security systems detect crowdturfers?''. Specifically, can Twitter's security system detect the artificial followers that were used for crowdsourced manipulation? To answer this question, we checked how many of our new followers were suspended by the Twitter Safety team two months after we collected them through Fiverr. We accessed each follower's Twitter profile page by using Twitter API. If the follower had been suspended by Twitter security system, the API returned the following error message: ``The account was suspended because of abnormal and suspicious behaviors''. Surprisingly, only 11,358 (24.6\%) of the 46,176 followers were suspended after two months. This indicates that Twitter's security system is not effectively detecting these manipulative followers (a.k.a. crowdturfers). This fact confirms that the web ecosystem and services need our crowdturfing task detection system to detect crowdturfing tasks and reduce the impact of these tasks on other web sites.



\section{Related Work}
In this section, we introduce some crowdsourcing research work which focused on understanding workers' demographic information, filtering low quality answers and spammers, and analyzing crowdturfing tasks and market.

Ross et al. \cite{Ross:2010} analyzed user demographics on Amazon Mechanical Turk, and found that the number of non-US workers has been increased, especially led by Indian workers who were mostly young, well-educated males. Heymann and Garcia-Molina \cite{Heymann:2011} proposed a novel analytics tool for crowdsourcing systems to gather logging events such as workers' location and used browser type.

Other researchers studied how to control quality of crowdsourced work, aiming at getting high quality results and filtering spammers who produce low quality answers. Venetis and Garcia-Molina \cite{Venetis:2012} compared various low quality answer filtering approaches such as gold standard, plurality and work time, found that the more number of workers participated in a task, the better result was produced. Halpin and Blanco \cite{halpin2012machine} used a machine learning technique to detect spammers at Amazon Mechanical Truck.

Researchers began studying crowdturfing problems and market. Motoyama et al. \cite{Motoyama:2011} analyzed abusive tasks on Freelancer. Wang et al. \cite{Wang:2012} analyzed two Chinese crowdsourcing sites and estimated that ~90\% of all tasks were crowdturfing tasks. Lee et al. \cite{Lee13icwsm} analyzed three Western crowdsourcing sites (e.g., Microworkers.com, ShortTask.com and Rapidworkers.com) and found that mainly targeted systems were online social networks (56\%) and search engines (33\%). Recently, Stringhini et al. \cite{stringhini:follower} and Thomas et al. \cite{thomas2013trafficking} studied Twitter follower market and Twitter account market, respectively.

Compared with the previous research work, we collected a large number of active tasks in Fiverr and analyzed crowdturfing tasks among them. We then developed crowdturfing task detection classifiers for the first time and effectively detected crowdturfing tasks. We measured the impact of these crowdturfing tasks in Twitter. This research will complement the existing research work.

\section{Conclusion}
In this paper, we have presented a comprehensive analysis of gigs and users in Fiverr and identified three types of crowdturfing gigs: social media targeting gigs, search engine targeting gigs and user traffic targeting gigs. Based on this analysis, we proposed and developed statistical classification models to automatically differentiate between legitimate gigs and crowdturfing gigs, and we provided the first study to detect crowdturfing tasks automatically. Our experimental results show that these models can effectively detect crowdturfing gigs with an accuracy rate of 97.35\%. Using these classification models, we identified 19,904 crowdturfing gigs in Fiverr, and we found that 70.7\% were social media targeting gigs, 27.3\% were search engine targeting gigs, and 2\% were user traffic targeting gigs. Then, we presented detailed case studies that identified important characteristics for each of these three types of crowdturfing gigs.

Finally, we measured the real world impact of crowdturfing by purchasing active Fiverr crowdturfing gigs that targeted Twitter. The purchased gigs generated tens of thousands of artificial followers for our Twitter accounts.  Our experimental results show that these crowdturfing gigs have a tangible impact on a real system. Specifically, our Twitter accounts were able to attain increased (and undeserved) influence on Twitter. We also tested Twitter's existing security system to measure its ability to detect and remove the artificial followers we obtained through crowdturfing. Surprisingly, after two months, the Twitter Safety team was only able to successfully detect 25\% of the artificial followers. This experimental result illustrates the importance of our crowdturfing gig detection study and the necessity of our crowdturfing detection classifiers for detecting and preventing crowdturfing tasks. Ultimately, we hope to widely deploy our system and reduce the impact of these tasks to other sites in advance.

\section{ Acknowledgments}
This work was supported in part by Google Faculty Research Award, faculty startup funds from Utah State University, and AFOSR Grant FA9550-12-1-0363. Any opinions, findings and conclusions or recommendations expressed in this material are the author(s) and do not necessarily reflect those of the sponsor.

{\small	
\bibliographystyle{aaai}
\bibliography{www}
}
\end{document}